\documentclass[10pt,aps,prd,amsmath,twocolumn,longbibliography]{revtex4-1}

\usepackage{amsmath}
\usepackage{graphicx}
\usepackage{amssymb,color}
\usepackage[utf8]{inputenc}
\usepackage[T1]{fontenc}
\usepackage{array}
\newcolumntype{P}[1]{>{\centering\arraybackslash}p{#1}}
\def\Dslash{\raise.15ex\hbox{/}\kern-.7em D}
\def\Pslash{\raise.15ex\hbox{/}\kern-.7em P}

\newcommand{\la}{\langle}
\newcommand{\ra}{\rangle}
\newcommand{\bfr}{{\bf r}}

\newcommand{\ben}{\begin{displaymath}}
\newcommand{\een}{\end{displaymath}}
\newcommand{\be}{\begin{equation}}
\newcommand{\ee}{\end{equation}}
\newcommand{\bea}{\begin{eqnarray}}
\newcommand{\eea}{\end{eqnarray}}

    \newcommand {\boldsigma}{\mbox{\boldmath$\sigma$}}

\newcommand {\boldnabla}{\mbox{\boldmath$\nabla$}}

\def\r{\rho}\def\L{\Lambda}\def\a{\alpha}

\def\m{\mu}

\newcommand{\beqn}{\begin{equation}}
\newcommand{\eeqn}{\end{equation}}

\usepackage{textpos}

\begin{document}                                           

\begin{textblock*}{100mm}(.3\textwidth,-2cm)
NT@UW-26-2
\end{textblock*}
 
\title{Proton-Size Resolution of the Hyperfine Puzzle in Hydrogen
}

\author{   Gerald A. Miller$^1$ }

\affiliation{$^1$ 
Department of Physics,
University of Washington, Seattle, WA 98195-1560, USA}
                                                                           
\date{\today}     
\begin{abstract}
Baym and Farrar (arXiv:2601.02300v1) have recently pointed out a puzzle in understanding the role of the  hyperfine interaction in the ground state of a hydrogen atom.  If one uses a variational wave function in which the Bohr radius, $a_0$ is replaced by a variational radius parameter,  $R$, first-order perturbation theory can give a contribution to the energy  proportional to $-1/R^3$. This raises the question of why the hyperfine interaction does not lead to collapse of
hydrogen. I show that including the effects of the non-zero size of the proton leads to a resolution of the puzzle such that the variational procedure yields a value of $R$ that is indistinguishable from $a_0$. 
\end{abstract}
\maketitle

\section{Introduction}
  
  The hyperfine interaction in hydrogen is a perturbation of order $\a^2 {m_e\over m_p}$ that splits the energy levels of different total (electron plus proton)  spin states~\cite{Eides:2007exa}. In a singlet state the hyperfine interaction is attractive, and neglecting the non-zero size of the proton, causes an interaction proportional to a three-dimensional Dirac delta function
so that its expectation value is $\propto 1/R^3$, where $R$ is a parameter characterizes  the size of the system.    

Baym \& Farrar~\cite{Baym:2026kqm} show that if one takes a variational approach to the hydrogen atom, with
the wave function given by
\bea 
\phi_R(r)={e^{-r/R}\over \sqrt{\pi R^3}},
\eea
a contribution to the energy for the spin-singlet  state, $E_{\rm hf,0}= -8{|\mu_e|\m_p\over R^3}$, results.
Here $\m_{e,p}$ are the magnetic moments of the electron and proton. This causes the energy to diverge to  $-\infty$ as the value of $R$ approaches 0.   Ref.~\cite{Baym:2026kqm} further explains that for values of $R$ smaller than the size of the proton the hyperfine interaction overwhelms the contribution of the kinetic energy and the system collapses. A  resolution is proposed that uses  the relativistic effects of the  Dirac equation. In that approach the magnetic moment of the electron depends on the variational parameter, $R$.    

I discuss the energy as a function of $R$ and propose a resolution based on the non-zero size of the proton. The wave function $\phi_R(r)$ is the exact solution for the non-relativistic hydrogen ground state when $R=a_0$, the Bohr radius. The kinetic energy for this wave function is  $1/(2m R^2)$ where $m$ is the electron mass, and the Coulomb energy is $-\a/R$, with $\a$ the fine structure constant. The difference between the electron mass and the reduced mass, $mM/(m+M)$  ($M$ is the proton mass) is ignored here and in Ref.~\cite{Baym:2026kqm}. Neglecting the hyperfine interaction, the expectation value of the energy is given by
\bea
E_0(R)={1\over 2m R^2} -{\a\over R}
.\eea 
This quantity has a  minimum of $E_0=-m\a^2/2$  at  $R=1/m\a=a_0$.

The Hamiltonian of the electron-proton hyperfine interaction between the electron and the proton is given by
\bea 
H_{\rm hf}(r)=|\m_e|\boldsigma_e\cdot {\bf H}(r),
\eea
where  $|\m_e|=e/(2m)$ and $\boldsigma_e$ is the electron Pauli matrix. The operator ${\bf H}(r)$ is given by an equation on page 58 of Bjorken and Drell~\cite{Bjorken:1965sts}:
\bea
{\bf H}(r)=-{g_p\over 2M}\int d^3r' \r(r') \boldnabla \times (\boldsigma_p\times \boldnabla) {1\over 4\pi |\bfr-\bfr'|},\nonumber\\
\eea
 with $\r(r')$ the magnetic moment density of the proton. This function takes the non-zero size of the proton into account.
 The proton gyromagnetic ratio, $g_p=2.79$ 
 Using the relation $ \boldnabla \times (\boldsigma_p\times \boldnabla)=(\boldsigma_p \nabla^2-\boldnabla(\boldsigma\cdot\boldnabla)$ and taking the angular average for spherically symmetric wave functions,
 one finds
 \bea
 & {\bf H}(r)=-{2\over3}{g_pe\boldsigma_p\over 2M}\int d^3r' \r(r') \nabla^2 {1\over 4\pi |\bfr-\bfr'|}\nonumber\\&={2\over3}{g_pe\boldsigma_p\over 2M}  \r(r).
\eea
The density $\r(r)$ is the three-dimensional Fourier transform of the Sachs magnetic form factor $G_M/\m_p$ when  computing the hyperfine interaction. I use the well-known dipole form
\bea
G_D(\vec q\,^2)={G_M(\vec q\,^2)\over\m_p} ={1\over (1+{\vec q\,^2\over \L^2})^2},
\eea
with $\L=0.843\, {\rm GeV}$.
This form is sufficiently accurate for the present purpose of showing that including a non-zero proton size is sufficient to resolve the hyperfine puzzle. Note that  the usual scalar argument $-q^2$ is here taken as $\vec q\,^2$ because the energy transfer in the one-photon exchange interaction may be neglected. The most recent version of $G_M$ is given in Ref.~\cite{Hague:2025gpe}. 
The  Fourier transform of $G_D$ is then obtained to be
\bea
\r(r)={\L^3\over 8\pi}e^{-\L r}
.\eea
This function approaches a three-dimensional delta function in the limit  $\L\to\infty$. If so, the expression for ${\bf H}$ reduces to that of Ref.~\cite{Baym:2026kqm}. 
One might think that the momentum scale of 0.843 GeV is irrelevant for the ground-state  energy of the hydrogen atom
because $\L =2.26 \times 10^5/a_0$.
 This is not the case, as is next  demonstrated.  
 
  The expectation value of the hyperfine integral is 
 \bea\la \phi_R| {\bf H }|\phi_R\ra={8e\over3}{g_p\boldsigma_p\over 2M}{1\over R^3} {\L^4R^3\over (2 +\L R)^3},
  \eea
and the  resulting  contribution of the hyperfine interaction to the energy of the hydrogen atom is  
\bea
E_{\rm hf}(R)={8\over3}{g_p\pi\a\over m M}{1\over R^3} {\L^3R^3\over (2 +\L R)^3} \la \boldsigma_e\cdot\boldsigma_p\ra.
\eea
This expression also reduces  that of Ref.~~\cite{Baym:2026kqm} in the limit that $\L\to\infty,$ and importantly is finite-valued   at $R=0$. The total energy, $E_1$ is now the sum of two terms:
\bea
E_1(R)=E_0(R)+E_{\rm hf}(R)
.\eea

\begin{figure}[h]
  \centering
 \includegraphics[width=0.3\textwidth]{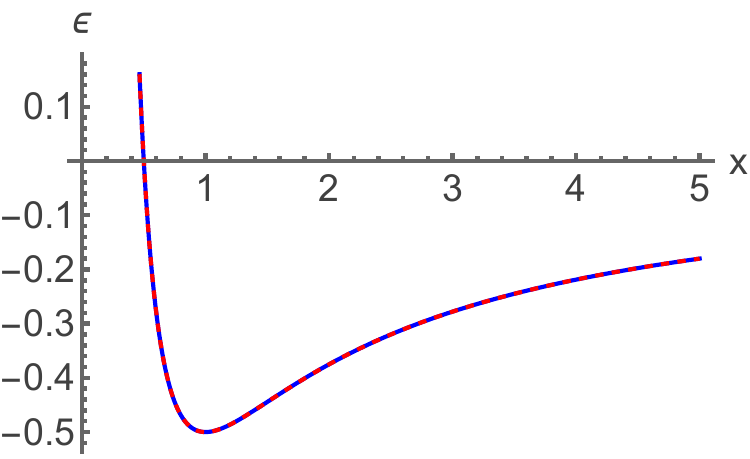}
  \includegraphics[width=0.3\textwidth]{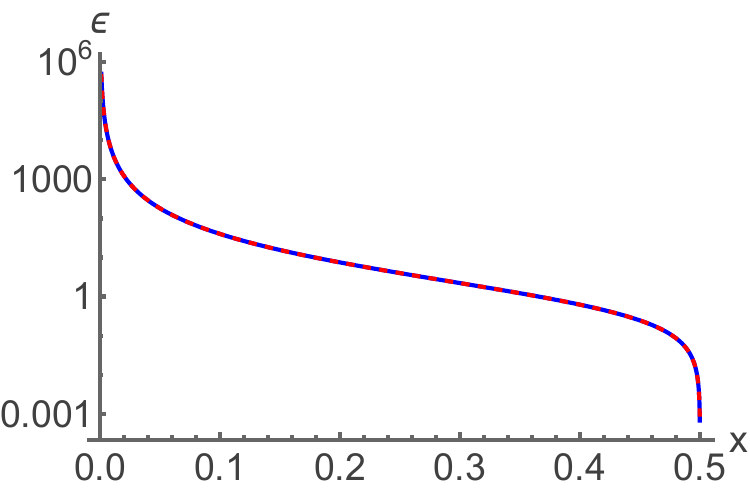}
  \caption{ $\epsilon_0(x)$  solid,  $\epsilon_1(x)$ dashed. The curves overlap.}
\end{figure}

The next step is to assess the influence of the term $E_{\rm hf}(R)$.  This is done using two methods. The first is an estimate using closed-form expressions. The second is numerical evaluation. The estimate is obtained by expanding $E(R)$ about the point $R=a_0$. This gives:
\begin{widetext}
\bea
E_1(R)\approx E_0(a_0)+{1\over 2}(R-a_0)^2 E^{''}_0(a_0) +E_{\rm hf}(a_0) +(R-a_0)E^{'}_{\rm hf}(a_0).
\eea
\end{widetext}
The using $\L a_0\gg 2$, the minimum of $E(R)$ occurs at 
\bea
R-a_0\approx- {E^{'}_{\rm hf}(a_0)\over  E^{''}_0(a_0) }= 24\pi\a g_p{1\over M},
\eea
for the spin-singlet state. This shift in the value of $R$  at the minimum of energy proportional to  the proton Compton wave length and is  6 $\times 10^{-6}\,a_0$. This is a truly negligible change in the value of $R$ at which $E(R)$ is minimized. 

The next step is to display the energy as a function of the variational parameter.  The hugely different momentum scales ($\L,1/a_0$)  makes it efficient to  use a rescaled variable $x=R/a_0$ and study the dimension-less quantities
$\epsilon_i =a_0/\a E_i(x)$. The results are shown in Figs.~1. The minima occur, as expected, at $x=1$. There is essentially no difference between the curves    for $\epsilon_0$  and  $\epsilon_1$, as shown in Fig.~1.   The upper part displays results for $x$ greater than about $x=0.4$. A log plot for smaller values of $x$ is also shown to make it clear that there is no other minimum than the one around $x=1$.

The results shown in Fig.~1 do not depend on the details of the magnetic form factor. Any form of $\r(r)$,  controlled by the proton size, reducing to a three-dimensional delta function in the limit that the size parameter vanishes, would give results nearly identical to those of Fig.~1.

In conclusion, including the effects of the non-zero size of the proton causes the effects of the hyperfine interaction in the singlet state to have virtually no impact on the energy of the state for all values of the variational parameter, $R$.
\begin{widetext}
\section*{Acknowledgements}
This work is partially funded by the  U.S. Department of Energy, Office of Science, Office of Nuclear Physics, under Contract No. DE-SC0026252. 
\end{widetext}
%

 
\end{document}